\documentclass[12pt]{article}
\usepackage{amsmath,amsfonts,amssymb}

\newcommand{\mathsym}[1]{{}}
\def\id{\protect{{1 \kern-.28em {\rm l}}}}

\def\be{\begin{eqnarray}}
\def\ee{\end{eqnarray}}
\def\p{{\partial}}

\makeatletter
\renewcommand\section{\@startsection {section}{1}{\z@}%
                                   {-3.5ex \@plus -1ex \@minus -.2ex}%
                                   {2.3ex \@plus.2ex}%
                                   {\normalfont\large\bfseries}}
\renewcommand\subsection{\@startsection{subsection}{2}{\z@}%
                                   {-3.25ex\@plus -1ex \@minus -.2ex}%
                                   {1.5ex \@plus .2ex}%
                                   {\normalfont\normalsize\bfseries}}

\makeatother

\def\b{{\rm b}} 
\def\d{{\rm d}} 

\def \foot {\footnote}
\def \bi{\bibitem}
\def \tr {{\rm tr}}
\def \ha {{1 \over 2}}

\def \ci{\cite}
\def \N {{\mathcal N}}
\def \ww {\Omega}

\def \XX {{\rm X}}
\def \cN {{\cal N}}

\def \d {\del}

\def\a{\alpha}
\def\b{\beta}

\def\p{\phi}

\def\P{{\bf P}}

\def \del{\partial}
\def \a {\alpha}

\def\g{\gamma}
\def\ov{\over}

\def\b{\beta}
\def\l{\lambda}

\def\foot{\footnote}

\def \ci {\cite}

\def \foot {\footnote}
\def \bi{\bibitem}
\def \ha {{1 \over 2}}

\def \XX {{\rm X}}

\def \d {\partial}

\def \P {\Phi}
\def \l  {\lambda}

\def \N {{\mathcal N}}

\def \N {{\mathcal N}}

\def \m {\mu}

\def \cD {{\cal D}}

\def \bi{\bibitem}
\def \la {\label}

\def \l {\lambda}
\def\foot{\footnote}

\newcommand{\rf}[1]{(\ref{#1})}
\def \ov {\over}

\def\N{{\cal N}}

\def \ha{{1\ov 2}}

\def \no {\nonumber}

\def \del {\partial}

\def \cN {{\cal N}}

\def \bi{\bibitem}
\def \la {\label}

\def \l {\lambda}
\def\foot{\footnote}

\def \HH {{\rm E}}

 \def \p {\phi}

\def \ov {\over}

\def \varpi {{\rm w}}

\def \d {\delta}

\def \bea {\be}
\def \eea {\ee}
  \def \d {\delta}

\def \DD {{\rm D}}

\def \b {\beta}

\def \del {\partial}
\def \d {\partial}
\def \s {\sigma}

\def \p {\phi}

\def \d {\del}

\def \ww  {\omega}

 \def \n {\nu}

\def \vp {\varphi}

\def \ed {\end{document}}
  
\def \cC {{\cal C}}

\def \ha {{{\textstyle{1 \ov 2}}}}

\def \vp {\varphi}

\def\a{\alpha}
\def\b{\beta}
\def\p{\partial}

\def\tr{{\rm tr}\,}

\def\cN{{\cal N}}
\def\cD{{\cal D}}

\def\bea{\begin{eqnarray}}
\def\eea{\end{eqnarray}}

\def\cN{{\cal N}}
\def\cC{{\cal C}}
\def\f{\frac}
\def\n{\nabla}
\def\tr{{\rm tr}\,}

\def\s{\sigma}

\def\d{\delta}
\def\q{\quad}
\def\g{\gamma}
\def\ci{\cite}
\def\l{\label}

\begin{document}


\overfullrule=0pt
\parskip=2pt
\parindent=12pt
\headheight=0in \headsep=0in \topmargin=0in \oddsidemargin=0in

\vspace{ -3cm}
\thispagestyle{empty}
\vspace{-1cm}

\rightline{Imperial-TP-AAT-2012-04}


\begin{center}
\vspace{1cm}
{\Large\bf
``Induced''  $\N=4$   conformal supergravity

\vspace{1.2cm}

  }

\vspace{.2cm}
 { I.L. Buchbinder$^{a}$,  N.G. Pletnev$^{b}$
and
 A.A. Tseytlin$^{c,}$\footnote{Also at Lebedev  Institute, Moscow.}
}
\vskip 0.6cm
{
\em
\vskip 0.08cm
\vskip 0.08cm
$^{a}$Department of Theoretical Physics, Tomsk State Pedagogical University,\\ Tomsk, 634061 Russia
\vskip 0.08cm
\vskip 0.08cm
$^{b}$Department of Theoretical Physics, Institute of Mathematics, \\ Novosibirsk, 630090 Russia
\vskip 0.08cm \vskip 0.08cm
 $^{c}$Blackett Laboratory, Imperial
College, London SW7 2AZ, U.K.
 }
\vspace{.2cm}
\end{center}

\begin{abstract}
We consider an Abelian $\N=4$  super Yang-Mills  theory   coupled to
background  $\N=4$ conformal supergravity fields. 
At the   classical level, this coupling
is invariant under global $SU(1,1)$ transformation of
the complex (``dilaton-axion'')  supergravity scalar
combined with an on-shell $\N=4$   vector-vector duality.
We compute the  divergent part of  the corresponding  
quantum effective action found by integrating over the
super Yang-Mills   fields and demonstrate its $SU(1,1)$ invariance. 
This divergent part   related to the  conformal anomaly is one-loop exact and
should be given by the  $\N=4$ conformal supergravity action containing
 the Weyl tensor squared term.
This allows us to  determine the full non-linear  form 
of the bosonic part of the  $\N=4$ conformal supergravity action
which  has manifest  $SU(1,1)$ invariance.

\end{abstract}

\newpage
\setcounter{equation}{0}
\setcounter{footnote}{0}
\setcounter{section}{0}
\renewcommand{\theequation}{1.\arabic{equation}}

\def \gv   {{\rm g}}
\def \DD {{\rm D}}

\def \HH {{\cal H}}  \def \SS {{\rm S}}
\def \XX {{}}

\def \p {\phi}
\def\bC {{\zeta}}
\def \P  {\p}
\def \ha {{\textstyle {1 \ov 2}}}
\def \trd {{\textstyle {3 \ov 2}}}
\def \Im {{\rm Im\,}}

\def \ww {{\rm w}}

\def \N {{\cal N}}
\def\pphi  {{\varphi}}
\def \vp {\sigma}


\section{Introduction}

The $\N=4$ conformal supergravity
(CSG)  as formulated in  \ci{bdw}    should have   global  $SU(1,1)$
or $SL(2,R)$
 symmetry   acting on  the singlet
complex scalar  (described by a 4-derivative analog of  the $SU(1,1)/U(1)$ coset
 sigma model).\footnote{To make this symmetry linearly realized  one may  introduce
also a spurious  local $U(1)$ symmetry.}
While the complete $\N=4$ superconformal 
  transformation laws  were written down in \ci{bdw},   the 
  full  non-linear action of such $\cN=4$
 conformal supergravity  
 was not explicitly  constructed  so far. 
 The aim   of this paper  is to find the  full bosonic part of such action.

 This  manifest $SU(1,1)$  symmetry is in
 general  broken  if one
  couples  the  $\cN=4$ CSG to $\cN=4$  super Yang-Mills (SYM)   theory
 \ci{roo,rw}. It is,  however,   preserved  in an  weaker  ``on-shell''
  form  in the case when  the   $\cN=4$ SYM  theory is {\it abelian}:
  the resulting
{\it equations of motion} are invariant under the $SU(1,1)$  acting
not only  on the complex scalar but also on the Abelian SYM vector
via
 vector-vector duality transformation.\footnote{This on-shell symmetry can
  be promoted to a manifest
symmetry of the action (at the expense of manifest Lorentz symmetry)
 if one uses a phase-space type  formulation
where one  doubles the  number of  vectors, see, e.g., \ci{ss}.}
 This symmetry is  then inherited by
the equations   of motion of the $\cN=4$ Poincare  supergravity  \ci{fer}
 as
it  can   be  obtained \ci{roo}
  from a system of 6  abelian vector
multiplets coupled to the $\cN=4$ conformal
supergravity multiplet.\foot{This  can be done by partial gauge fixing
and solving for some of the CSG fields that in the absence of the pure CSG action
 play a role of  auxiliary fields \ci{roo,rw}. Potential importance
 of  superconformal
 formulation of $\cN=4$ Poincare  supergravity was recently
 emphasised  in \ci{ren}.} 

As was found in \ci{ft1,ft2},
the  $SU(1,1)$ invariant    $\cN=4$ CSG of  \ci{bdw}  has non-zero
beta-function or  conformal anomaly  and is thus  inconsistent at the
quantum  level unless it is  coupled to   {\it four}
  $\cN=4$  vector  multiplets  (see \ci{ftr} for a review).
   This
conclusion was confirmed in \ci{rh} on the basis of
analysis of the  local $SU(4)$    chiral  anomaly
 (which is in the same  multiplet  with  trace anomaly).

At the same time, it was suggested in \ci{ft1,ft2} that there  might
exist an alternative version of $\cN=4$ CSG   without the  $SU(1,1)$
invariance in which a non-minimal   coupling 
of the
singlet scalar to the square of the Weyl tensor  may be  present. For
a particular   value of such coupling the resulting ``non-minimal''
$\cN=4$ CSG  can be made UV finite by itself, i.e. without adding   extra $\cN=4$ vector
multiplets  \ci{ft1}.\foot{It is not clear, however, how this
conjecture can be reconciled with  the $SU(4)$ anomaly cancellation study \ci{rh}
which does not seem to be
sensitive to such  non-minimal terms. That suggests a
potential problem with realization of supersymmetry  which  should be
requiring that  all superconformal anomalies should belong to one  supermultiplet.}
Curiously,   a  similar type of  ``non-minimal''
  $\cN=4$  conformal supergravity
seems  to  emerge   \ci{bw}   in the twistor-string  \ci{w}  context.

The coupling between $\cN=4$ SYM and $\cN=4$ CSG multiplets
  appears also in the context
of  the AdS/CFT correspondence \ci{ff,lt,t}: the $\cN=4$  SYM path
integral with the CSG  fields  as external ``sources''  may
be interpreted   as a generating functional for correlators of  particular 1/2
BPS operators (dimension 2  chiral primary  operator and its
  supersymmetry  descendants, i.e. the  fields of the
stress  tensor  multiplet   dual to $\cN=8,d=5$ supergravity fields).
 After integrating over the quantum SYM fields, the
  conformal supergravity action should then be
  the coefficient of the   logarithmic divergence   in the
 resulting effective action. In that limited sense the $\cN=4$ CSG   may be
   interpreted as an {\it ``induced''}  theory.\foot{The full SYM
  effective action in CSG background
   contains of course also  a finite non-local part, see \ci{lt}.
   While the divergent part will preserve all the classical superconformal
    symmetries,
   the finite non-local part will  contain non-invariant anomalous terms.}

 Since the  superconformal   anomaly  should be 1-loop exact,
 the result for the logarithmic divergence
 should   be   given    just  by  the 1-loop  contribution.\footnote{It is
  thus the  same at weak  and at  strong SYM coupling  and  can be
  also  found  by
  evaluating the  $d=5$  supergravity   action on the solution of
  the corresponding Dirichlet problem (from the cutoff-dependent
   part of the resulting expression \ci{lt}).}
This also  means   that the  divergent term
 is not sensitive to the
 non-Abelian structure of the SYM
  theory, i.e. it is sufficient to consider just  {\it one} abelian
 $\cN=4$  vector multiplet coupled to the
 external  $\cN=4$ CSG multiplet and do  the gaussian integral
   over the $\cN=4$ vector   multiplet fields.

   As  the full non-linear form of the coupling
   between  the  $\cN=4$  SYM and CSG
    multiplets is known
   \ci{roo,rw},  and since the  one-loop logarithmic divergence of the   $\cN=4$
   vector multiplet
   fields  is determined  by a relevant
    Seeley coefficient of
     the corresponding
   2nd order matrix differential operator (with coefficients depending on
   the  external CSG fields)
   it   should  thus  be  straightforward to reconstruct  the
   full non-linear form of the
   resulting $\cN=4$ CSG action  using the standard algorithm \ci{gilk}, i.e.  one should
   get \ci{lt}
   \bea
  && \Gamma_\infty = - (\ln Z_{_{{\rm \cN=4\, SYM}}})_{\infty}
    = \  k  \ I_{_{ {\rm \cN=4\, CSG}} }\ , \ \ \ \ \ \ \ \ \ \ \ \ \
    k=  - {N^2 \ov 4 (4 \pi)^2} \ln \Lambda \ , \la{cs}\\
    && I_{_{\rm  \cN=4\, CSG} }= \int d^4 x \sqrt g\   {\cal L}_{_{ {\rm \cN=4\, CSG}} } =    \int d^4 x \sqrt g\ ( C^2 + ...)    \ , \la{ccc}
     \eea
    where $N$ is the number of $\cN=4$ vector multiplets,  $\Lambda$ is a UV cutoff.
    Here  $I_{\rm \cN=4\, CSG}$   
    should be  the   CSG action as it   starts with the Weyl tensor squared $C^2$ term
    (up to  total derivative   Euler density term):
   since  $I_{\rm \cN=4\, CSG}$  should inherit  all the symmetries of $\cN=4$ conformal 
    supergravity by construction\foot{
   The invariance of the divergent part can be seen  explicitly if one uses, e.g., 
   dimensional   regularization. 
   Let ${\Gamma}_{\rm reg} =   \frac{1}{n-4}  {\Gamma}_{\rm div} + {\Gamma}_{\rm fin}   $
    be  the regularized effective
   action. 
   Then under a superconformal transformation 
    ${\delta}{\Gamma}_{\rm reg}$ = $(n-4){\cal A} $,   so that 
    ${\delta}{\Gamma}_{\rm div}$= $ 0 $  and 
    ${\delta}{\Gamma}_{\rm fin}$= $ {\cal A} $ (see, e.g., \ci{B} for details).}  and 
    contains the $C^2$ term it must  represent 
    the complete non-linear action of $\cN=4$
    conformal supergravity.

   In  particular,  since the coupling  between an  Abelian $\cN=4$ SYM and $\cN=4$ CSG  multiplets
    preserves
   the scalar $SU(1,1)$
   symmetry  combined with a duality rotation of the $\cN=4$ SYM  vector \ci{roo}
   and since the  latter   is integrated over in the path integral,
   the resulting ``induced''  CSG   action
   should have  manifest (off-shell) $SU(1,1)$  symmetry.\footnote{This
    follows, e.g., from the fact
   that the vector-vector duality   may be performed  as a change of variables in the path
   integral (in full analogy with 2d scalar-scalar or T-duality). More precisely,
   while the logarithmically divergent part of the path integral  should be invariant
     its
   finite part may  contain   a  local term  not invariant under the  $SU(1,1)$,
    similarly to what happens in the 2d   case  where the dilaton shifts  under  the T-duality
   (see \ci{rt} and refs. there).} 
   This was already  demonstrated in   \ci{os}  in  the  subsector   of
    the   standard  $SL(2,R)$  invariant scalar-vector  coupling
   ($ e^{- \vp} F_{mn}F_{mn} - i {\cal C} F_{mn}F^\star_{mn}$).
     Here we will
    demonstrate this  for the full $\cN=4$  vector -- CSG coupling case,
    thus determining  the full  $SU(1,1)$ invariant form of the bosonic part of
    the $\cN=4$ CSG  action.

This computation is of interest as the complete    non-linear form of the
     $\cN=4$ CSG action   was not  explicitly given     before.
   The   terms in the CSG action   which are quadratic in  the non-metric
   fields  (but non-linear   in the   metric)
   can be  reconstructed   \ci{ft1,ftr}   by requiring  the Weyl symmetry
   and  reparametrization
   invariance,  but  higher order terms
   are hard to determine directly.\footnote{In principle, they
    can  be reconstructed  using the
   Noether
   procedure   given that the full non-linear supersymmetry
    transformation rules are known (and close off shell on CSG fields) \ci{bdw}.}
  The non-linear terms of $\cN=4$
   CSG action should of course reduce   to the
   corresponding terms in the  full $\cN=2$ CSG
   action which was found in \ci{bdw}; this  provides a non-trivial check.

As the ``induced''  CSG action  we find  below
is manifestly $SU(1,1)$ invariant,
an  apparent   absence of an alternative  to
 the  $SU(1,1)$ invariant  coupling \ci{roo} between
the Abelian $\cN=4$  SYM   and $\cN=4$  CSG   multiplets
 appears to rule out   the possibility
of some     $SU(1,1)$ non-invariant  ``non-minimal''   conformal supergravity model.


We shall start in  section 2 with a review of  the Lagrangian of an
Abelian $\cN=4$  vector multiplet coupled to (bosonic part of)
$\cN=4$   conformal supergravity background. In section 3  we shall
compute the  UV divergent part of the
 effective
action found by integrating over  the vector multiplet  fields
and show that the resulting   $SU(1,1)$   invariant   expression
  has the expected structure of the
  $\cN=4$  CSG  action. A short summary will  be  given in section 4.

\renewcommand{\theequation}{2.\arabic{equation}}
 \setcounter{equation}{0}
\section{$\N=4$   Abelian vector multiplet   coupled
 to external $\N=4$ conformal supergravity}

Let us  start with a  review of  the  action \ci{roo} for  an
 Abelian $\N=4$  vector multiplet in a background of $\N=4$ conformal supergravity.
We shall denote the vector multiplet fields as ${\cal A}=\{A_m, \pphi_{ij}, \psi_i\}$.
In what follows  $m,n,r,s=1,2,3,4$ are space-time   indices and
$i,j,k,l=1,2,3,4$ are $SU(4)$   indices.
The scalar fields satisfy
the conditions
\be \la{re}
\pphi_{ij}=-\pphi_{ji}=-\f12\varepsilon_{ijkl}\pphi^{kl} \ , \ \ \ \ \ \ \ \ \ \ \ \
\pphi^{ij}=(\pphi_{ij})^\ast   \ . \ee
For the fermions
$\psi^i=P_+\psi^i$ transforms as
$\bf{4}$ of $SU(4)$, and  $\psi_i\equiv P_-\psi^i= (\psi^i)^\ast,\
 \bar\psi^i\equiv\bar\psi^iP_+,\ \bar\psi_i\equiv\bar\psi^i
P_-$, where $P_{\pm}$ are chiral projectors.

The bosonic CSG  fields  \ci{bdw}  are   ${\cal G}=\{e^a_m, V^{i}_{j \ m},
T^{-ij}_{mn}, \bC, E_{ij}, D^{ij}_{\ \ kl}\}$, while the  fermionic fields are
$\{\psi^i_m,
\Lambda_i,\chi^{ij}_{\ \ k}\}$.  In what follows we shall consider only the bosonic
CSG background.

Here $e^a_m$ is the vierbein, $ V^{i}_{j\, m}$ is   $SU(4)$ gauge field potential,
$T^{-ij}_{mn}$ are complex antisymmetric
antiselfdual tensors of dimension 1 transforming in $\bf{6}$ of $SU(4)$
($ T^{-ij}_{mn} =-\f12\varepsilon_{mn}^{\ \ pq}T^{-ij}_{pq} $)
while $(\bC, E_{ij}, D^{ij}_{\ \ kl})$
are  Lorentz scalars   of dimensions 0, 1 and 2 respectively
(i.e. they have 4, 2 and 0 derivatives in their kinetic term in CSG action \ci{bdw,ftr}).
The complex scalars $E_{ij}=E_{ji} $ are in representation $\bf{10}$ of $SU(4)$, while
$D^{ij}_{\ \ kl}$  are in real representation $\bf{20}$
($D^{ij}_{\ \ kl}=D_{kl}^{\ \ ij}=(D^{ij}_{\ \ kl})^\ast=\f14\varepsilon^{iji'j'}
\varepsilon_{klk'l'}D^{k'l'}_{\ \ i'j'})$.

In \ci{bdw} the physical complex  scalar  $\bC$ is replaced by a  doublet  of complex scalars
$\P_\a$ with
\be \P^\a\P_\a= \P_1 \P_1^*-\P_2 \P_2^*  =1  \ ,     \ \ \ \ \ \ \ \ \
 \phi^1=(\phi_1)^\ast\ , \ \ \ \phi^2=-(\phi_2)^\ast \ ,      \la{11} \ee
by adding a local $U(1)$ gauge symmetry. Then   $\P_\a$ transforms under global $SU(1,1)$
as well as local $U(1)$,
 $\P'_\a = e^{-i \g (x) }   U_{\a}^\b \P_\b $, i.e.
has the  $U(1)$ chiral weight $-1$.\foot{Other CSG fields   having  non-zero  chiral
 weights  are:   $T^{-}{}^{ij}_{mn}$ (-1); \   $E_{(ij)}$ (-1);
  $\Lambda_i\ (- \trd)$;
  $\chi_k^{[ij]}\ (- \ha)$;
 $\psi^i_\m\ (-\ha)$. The $Q$-susy parameter $\epsilon_i$  has weight 1/2.
}
 Then only $\P_\a$ transforms under $SU(1,1)$  but other fields   with non-zero
chiral  weights transform  under
 local $U(1)$, i.e.  all fields
   with derivative couplings and non-zero chiral weights
 couple to the  scalar  $U(1)$   connection
  through the  covariant derivative  ($\ww$ is the chiral weight)
 \be\la{1}
\DD_m   = \del_m    -  i \ww a_m   \ , \ \ \ \ \ \ \ \ \ \
      a_m=  { i } \P^\a \del_m \P_\a 
       \ .
 \ee
  The scalar connection $a_m$ is invariant  under the $SU(1,1)$  and transforms
  by a gradient  under the $U(1)$. 

  The general  form \ci{roo} of the $\N=4$ vector multiplet  Lagrangian
(before $U(1)$ gauge fixing) may be written as \ci{roo} ${\cal
L}={\cal L}_{B}+{\cal L}_{F}$,  with  the bosonic part\foot{We  use
Euclidean signature with imaginary  time (fourth) component, with
$\varepsilon^{1234} =1$. For simplicity we  shall  often ignore
trivial metric factors  not distinguishing between coordinate and
target-space indices (which are always
 contracted with Euclidean  signature metric so we will often not raise them in the  contractions).
 Self-dual parts of 2nd rank tensors are defined as
$
F^\pm_{mn}=\f12(F_{mn}\pm  F^\star_{mn})\ , \  F^+_{mn}= ( F^-_{mn})^* $,
$ F^\star_{mn}=\f{1}{2}\varepsilon_{mn pq}F^{pq}$.}
\bea\l{act}
&& {\cal L}_{B}=  \f14 i \tau(\phi) F^+_{mn}F^+_{mn}  
-\f14 i  \bar \tau(\phi)F^-_{mn}F^-_{mn}  
\no \\
&&\ \ \ \ \  \ \ -\big(\f{1}{\Phi}T^+_{mn ij}F^+_{mn} \pphi^{ij}
+\f{1}{\Phi^\ast}   T^{-ij}_{mn}F^-_{mn} \pphi_{ij}  \big) -\f12\big(\f{\Phi^\ast}{\Phi}
T^+_{mn ij}T^+_{mn kl}\pphi^{ij} \pphi^{kl}
 +\f{\Phi}{\Phi^\ast} T^{-ij}_{mn}T^{-kl}_{mn}\pphi_{ij}\pphi_{kl}  \big) \no\\
&&\ \ \ \ \ \ \
-\f12\DD_m\pphi^{ij}\DD_m\pphi_{ij}
-\f{1}{12}\big(R + \ha E^{kl}E_{kl}+2\DD_m\phi^\a \DD_m\phi_\a \big) \pphi^{ij}\pphi_{ij}
+\f{1}{4}D^{\  kl}_{ ij}\pphi_{kl} \pphi^{ij} \ , \la{ab}\\
&&  \la{222}  i  \tau(\phi) \equiv  - {\p^*_1 + \p^*_2 \ov \p^*_1 - \p^*_2} \ , \ \ \ \ \
i \bar \tau(\phi) =  {\p_1 + \p_2 \ov \p_1 - \p_2} \ , \ \ \ \ \
\Phi(\phi) \equiv  \p^*_1 - \p^*_2 \ , \ \ \ \ \ \Phi^*= \p_1 - \p_2 \ ,
\eea
and the  fermionic part
\bea\l{act fermi}
&& {\cal L}_{F}=-\f12\bar\psi^i\not\!\!\DD\psi_i-\f12\bar\psi_i\not\!\!\DD\psi^i-\f14E_{ij}\bar\psi^i\psi^j-
\f14E^{ij}\bar\psi_i\psi_j \no\\
&& \ \ \ \ \ \ \ \ +\f14\varepsilon_{iklj}\bar\psi^i\s_{mn}
T^{-kl}_{mn} \psi^j+ \f14\varepsilon^{iklj}\bar\psi_i\s_{mn}
T^+_{mn \ kl}\psi_j \  . \la{f}\eea
In general, the  derivative 
$\DD_m$  contains the gravitational  $\n_m$ part
as well as 
 the $SU(4)$  gauge potential ($V_m$),
   in addition to the  $U(1)$   term ($a_m$) in \rf{1}
    (note that the bosonic  vector multiplet  fields
  have zero chiral weights while
  $\psi_i$ has weight -1/2).

While the  $F_{mn}(A)$ dependent part of the action \rf{ab}  is not invariant under  $SU(1,1)$  acting on $\phi_\a$,
it was shown in \ci{roo} that the corresponding equations of motion  (written in first order form)
are invariant provided one also ``duality-rotates''   the vector
field strength as in the closely related  case of  the
 Poincare supergravity \ci{fer}.

Our aim will be to integrate over  the vector multiplet fields
$\{A_m, \pphi_{ij}, \psi_i\}$  in \rf{ab},\rf{f}  and compute the divergent part of the
resulting effective action. For this we do not need
to fix the local $U(1)$ symmetry and may treat   the scalar
functions  $\tau(\phi),\, \Phi(\phi)$  and $a_m$ as arbitrary background fields.
Equivalently, we may  choose to fix  the
spurious   local $U(1)$  by  a ``physical'' gauge, e.g.,
 $\P_1 = \P_1^*$  \ci{bdw,roo}
  \be \la{ga}
  \P_1 = (1-  \bC\bC^*)^{-1/2}\ , \ \ \ \ \ \ \ \ \ \  \ \P_2 = \bC\,(1-  \bC\bC^*)^{-1/2} \ ,  \ee
  where the  complex  scalar $\bC$  (taking values in the disc $|\bC| \leq 1$)
   is an independent degree of freedom.
 Then $a_m$ is no longer a invariant  of a
 redefined $SU(1,1)$ acting on $\bC$  (that preserves the gauge condition)
 but it changes  only by a gradient.
  Explicitly,\foot{In our notation here $A_{[n} B_{m]}= A_n B_m - A_m B_n $.}
 \be
 a_m =  { i }   { \bC \del_m \bC^* -\bC^* \del_m \bC\ov 2 (1 -  \bC\bC^* )}  \ , \ \ \ \ \ \ \
 \ \ \ \ \ \
 F_{mn}(a) \equiv  \del_{[m}  a_{n]}=  {i }   { \del_{[m} \bC \del_{n]} \bC^* \ov (1 -  \bC\bC^*)^{2} }  \ .  \la{1.2} \ee
Instead of $\bC$   it is  useful to use  the   complex scalar   which is
directly equal
   to the scalar-vector coupling  $ \tau(\phi)$ in \rf{ab}
\be
&&\ \ \ \ \ \ \ \ \ \ \ \ \ \ \ \ \ \ \tau \equiv   \cC+ie^{-\vp}
=  i { \phi_1^* + \phi_2^* \ov \phi^*_1- \phi_2^*} =   i { 1 + \bC^* \ov 1- \bC^*} \ ,
\la{22}\\
&&a_m =-  { \del_m (\tau +\bar \tau)  \ov  4\, \Im \tau }  +
 \f i2 \del_m \ln\f{\tau+i}{\bar\tau-i} \ ,    \ \ \ \ \ \ \
F_{mn}(a) = i    \del_{[m} \phi^\a\del_{n]} \phi_\a =
 -    { \del_{[m} \tau \del_{n]} \bar \tau \ov 4  (\Im \tau )^{2} }  \ .
 \la{aaa}
\ee
The transformation   from $\bC$ to $\tau$ in \rf{22} maps a unit into half-plane,
so that $\tau$  transforms  as
  $\tau\to { a \tau + b \ov  c \tau + d}$ under the  corresponding  $SL(2,R)$
   equivalent to original $SU(1,1)$ (see, e.g., \ci{jo}).
One has  in \rf{222} \be \la{221}  i \bar \tau =  g^{-2} +
i  \cC    \ , \ \ \ \ \ \ \ \Phi\Phi^*= g^2 =(\Im \tau)^{-1} \ , \
\ \ \ \ \    g \equiv e^{\vp/2}\ .  \ee Note also that\foot{Here $\DD_m \phi_\a =
(\del_m + i a_m )\phi_\a$, see \rf{1}.
 $\DD_m \p^\a \DD_m \p_\a$   is manifestly $SU(1,1)$  invariant,
 and thus  invariant   under the $SL(2,R)$   acting on $\tau$, with
$\Im\tau\rightarrow \f{1}{(c\tau+d)(c\bar\tau+d)}\Im\tau,\q
\del_m\tau\rightarrow\f{1}{(c\tau+d)^2}\del_m\tau$.}
\be \la{3}
- 4 \DD_m \p^\a \DD_m \p_\a =  4 {\del_m \bC \del_m \bC^* \ov (1 -
\bC\bC^*)^2  } =  {\del_m \tau  \del_m \bar \tau  \ov (\Im \tau )^2}
= (\del_m \vp)^2 + e^{2 \vp} (\del_m \cC)^2
  \ .
\ee

\renewcommand{\theequation}{3.\arabic{equation}}
 \setcounter{equation}{0}
\section{Divergent part of  $\cN=4$  SYM
effective action in   conformal\\ supergravity background }

The UV divergent part  of the SYM effective action  in the CSG
background is related to conformal anomaly and thus should be given
to all orders  by the 1-loop logarithmically divergent   term. To
determine the  latter one  may just consider a single Abelian vector
multiplet action \rf{ab},\rf{f}  quadratic in ${\cal A}=\{A_m,
\pphi_{ij},  \psi_i\}$
 but keeping   full dependence on the (bosonic)  background fields
${\cal G}=\{e^a_m, V^{i}_{j  m}, T^{-ij}_{mn}, \bC, E_{ij}, D^{ij}_{\ \ kl}\}$.
As already mentioned,
while    it is not necessary  to  fix the $U(1)$ gauge
for concreteness we will be expressing  all the scalar   functions
  in terms of
the complex scalar  $\tau$ in \rf{22}--\rf{3}.

The 1-loop effective action is  given  by  the contribution of the
mixed vector-scalar sector, the  vector ghosts and the fermions
\be\label{det}\Gamma= \f12\ln \mbox{Det}{\cal H}_{1,0} - \ln
\mbox{Det}{\cal H}_{gh} - \f12\ln \mbox{Det}{\cal H}_{1/2}\ ,\ee
where ${\cal H}$ are second-order matrix
 differential operators, depending on the background fields ${\cal G}$.
Then
\be\la{gaa}
   \Gamma_\infty = -  {1 \ov (4\pi)^2} \ln \Lambda \int d^4 x \sqrt g \   (a_2)_ {\cN=4 \, \rm tot}   \ , \ee
 where  the  diagonal DeWitt-Seeley  coefficient  $a_2$ of the generic operator
 \be\l{cH} {\cal H}_{AB}=-1_{AB}\hat\nabla^2+2h^m_{AB}\hat\n_m+\Pi_{AB} \  \ee
 has   the following form \ci{gilk}
\bea\label{a2}
&& a_2= \mbox{tr}\Big[ \f{
1}{180}(R_{mn rs}R^{mnrs}-R_{mn}R^{mn}+\nabla^2 R)+\f{ 1}{6}\nabla^2
\hat{P}+\f12 \hat{P}\cdot
\hat{P}+\f{1}{12}\hat{\cal F}_{mn}\hat{\cal F}^{mn }\Big] \ , \no \\
&&  \hat{P}_{AB}=\Pi_{AB}-\f{1}{6} R\, 1_{AB}-\hat\n_m h^m_{AB}+h_{m AC} h^m_{CB} \ ,  \la{p}
\\&&
\hat{\cal F}_{mn AB}
=[\hat\n_m,\hat\n_n]_{AB}
-\hat\n_{[m}h_{n]AB}+h_{[m AC}h_{n]CB}   \ . \no
\eea
Here $\hat \n_m$  is given  by the gravitational covariant derivative  $\n_m$
 plus   possible
extra gauge ($SU(4)$ and $U(1)$)  field potentials   for unmixed fields,
while $h^m_{AB}$ accounts for the mixing between different types of fields.

The vector-scalar operator originating from
 from \rf{ab}  may be written as
\be\label{Hphi}
{\cal H}_{1,0}=\left(\begin{array}{ccc}{\cal H}_1 &
-2 g\overrightarrow{\DD_m}\f{1}{\Phi^\ast}T^{-kl}_{mn}
& -2 g\overrightarrow{\DD_m}\f{1}{\Phi}T^+_{mnkl}
\\2T^+_{ijnm}\f{1}{\Phi}\overrightarrow{\DD_n}g
& {\cal H}_0  & \f{\Phi^\ast}{\Phi}T^+_{ij} \cdot T^+_{kl}
\\2T^{-ij}_{nm}\f{1}{\Phi^\ast}\overrightarrow{\DD_n}g &
\f{\Phi}{\Phi^\ast}T^{-ij}_{} \cdot T^{-kl}_{}  & {\cal H}_0   \\ \end{array}\right) \ , \la{vs}
\ee
where $g= e^{\vp/2}$  is a  coupling function
(see \rf{221}), $\DD_m = \nabla_m   + i a_m -  V_m$   and
\be
 ({\cal H}_0)^{kl}_{ij} =\big(-\DD^2 +   \f16R +   \f{1}{12}M \big)1^{kl}_{ij}   -\f12D_{ij}^{\ \
kl} \ ,  \ \ \ \ \ \ \
 M=E^{kl}E_{kl}+4 \DD_m\phi^\a \DD_m\phi_\a    \ .  \la{0} \ee
The fermionic operator can be  found   by squaring the
first-order operator in \rf{f}
\be\label{Hpsi}-\f12\left(\begin{array}{cc}\bar\psi^i & \bar\psi_i
\\\end{array}\right)\left(\begin{array}{cc}\not\!\!\DD\d^j_iP_-
& (\f12E_{ij}+\s\cdot T^-_{ij})P_+
\\(\f12E^{ij}+\s\cdot T^{+ij})P_- & \not\!\!\DD\d_j^iP_+
\\\end{array}\right)\left(\begin{array}{c}\psi_j \\\psi^j \\\end{array}\right) \ .
\ee
Here
 $\DD_m=\del_m+\f12\s_{ab}\omega_m^{ab}  + { i \ov 2}  a_m-V_m$
 and $P_\pm $ are chiral projectors.

\subsection{Vector-scalar  sector}
\def \ne {\n}

Let us start  with  the contribution of the vector-scalar sector
 (in which   we  will include
also the ghost contribution).
Ignoring first the vector-scalar mixing  due to the $T^{-ij}_{mn}$ background
in \rf{ab}
 one  is to  account for   the presence of  a non-trivial  scalar background-dependent
factor in the vector  kinetic  operator $\HH_1$. This issue was  dealt
   with already
in \ci{os}  in the case of  a  simple  vector coupling in the first line of \rf{ab}
 and we will follow the same approach here.

Choosing the  gauge fixing term as
$ g^2[ \n_m(\f{1}{g^2}A_m)]^2 $ where $g=e^{\vp/2}$
and redefining $ A_m \to g A_m$  the  vector
operator ${\cal H}_1$  may be written as (here ${\cal C}$ is the real part of $\tau$ in
 \rf{22})
\bea && {\cal H}_{1mn} =
 g_{mn}(- \tilde\n^2 +\Pi) +  \Pi_{mn} \ , \la{ve}\\
 && \Pi_{mn}=R_{mn}+ g^4\n_m\f{1}{g^2}\n_n\f{1}{g^2}-g^2\n_m\n_n\f{1}{g^2} +
\f12g^4\big(g_{mn}\n_r{\cal C}\n_r{\cal C}-\n_m{\cal C}\n_n{\cal C}\big) \ ,
\cr
&&\Pi=\f12g^2 \n^2\f{1}{g^2} -\f14g^4 \n_m\f{1}{g^2}\ \n_m\f{1}{g^2} \ , \ \ \ \ \
\tilde\n_m
A_n\equiv \n_m A_n-\f{i}{2}g^2\varepsilon_{m n}^{\ \ \ rs}\n_r  {\cal C}A_s \ .  \la{pi}
\eea
The corresponding ghost operator  is 
\be  \la{gg} {\cal H}_{gh}=-\n^2 + \Pi  \ . \ee
 Then in addition to the   standard   single-vector gravitational contribution to
 $a_2$  \ci{du}\foot{We include the ghost contribution   and
  ignore the scheme-dependent total derivative term $\n^2 R$.}
 \bea
 && (a_2)_{1\, {\rm grav}} = \f{1}{10} C^2 -\f{31}{180} E  \ , \ \ \ \ \ \ \ \ \ \ \la{hq} \\
&& C^2= R^{mnpq}R_{mnpq}-2R^{mn}R_{mn}+\f13 R^2\ , \la{1g} \\
&&
 E\equiv  R^\star R^ \star =  R^{mnpq}R_{mnpq}-4R^{mn}R_{mn}+R^2  \ , \ \ \ \ \ \ \
\ \ \ C^2 - E =2 (R_{mn}^2-\f13R^2)\ , \no   \la{cccc}
 \eea
there is also a non-trivial   scalar  background  contribution \ci{os}
($\ne_m\tau=\del_m
\tau$)
\bea\la{sss}
&& {\SS}(\tau)=\f{1}{4(\Im\tau)^2}\Big[\cD^2\tau
\cD^2\bar\tau-2(R_{mn}-   \f13 R g_{mn})  \ne_m\tau\ne_n\bar\tau \Big]
\cr
&&
\ \ \ \ \ \ \ \ \  +\f{1}{48(\Im\tau)^4}\Big( \ne_m\tau\ne_m\tau\ne_n\bar\tau\ne_n\bar\tau+
2    \n_m\tau\n_m\bar\tau   \ne_n\tau\ne_n\bar\tau    \Big)  \ ,  \la{osi} \\
&& \ \ \ \cD^2\tau\equiv \nabla^2 \tau+\f{i}{\Im\tau}\ne_m\tau\ne_m\tau
,\ \ \ \
 \q \cD^2\bar\tau\equiv \nabla^2 \bar\tau-
\f{i}{\Im\tau}\ne_m\bar\tau\ne_m\bar\tau \ . \no \eea
The quadratic part
  of this  4-derivative action is the same as found for the singlet scalar
  kinetic term in the CSG action \ci{ftr}.
  The full non-linear   expression \rf{osi}  is invariant  under  the  $SL(2,R)$
     acting  on the local scalar coupling $\tau =  {\cal C}
 + i g^{-2}$  \ci{os} (note, e.g.,  that
 $\f{1}{\Im\tau}\cD^2\tau\rightarrow\f{c\bar\tau+d}{c\tau+d}\f{1}{\Im\tau}\cD^2\tau$).

 To compute  the scalar contribution
 we need to account for the  reality constraints \rf{re}: 
we may solve them explicitly\foot{A solution to these  constraints    may be chosen
as \be \no
\pphi_{ij}=\left(\begin{array}{cccc} 0 & \pphi_{12} & \pphi_{13} & \pphi_{14} \\
-\pphi_{12} & 0 & -\pphi_{14}^\ast & \pphi_{13}^\ast \\-\pphi_{13} &
\pphi_{14}^\ast & 0 & -\pphi_{12}^\ast\\-\pphi_{14}
 & -\pphi_{13}^\ast & \pphi_{12}^\ast & 0 \\\end{array}\right),  \ \ \ \ \
\q \del_m \pphi^{ij}   \del_m\pphi_{ij}=4(\del_m \pphi_{12}^\ast \del_m \pphi_{12}+ \del_m
\pphi_{13}^\ast  \del_m \pphi_{13}+\del_m   \pphi_{14}^\ast  \del_m \pphi_{14}).\ee} or formally
 do  the summation over $i,j$  in \rf{vs},  adding
extra 1/2 factor in the final result.

 The operator (\ref{vs}) has the  form (\ref{cH}) where
\bea   && 1_{AB}=\left(\begin{array}{ccc} g_{mn} & 0 & 0 \\0 &   1^{kl}_{ij} & 0 \\0 & 0 &
 1^{ij}_{kl} \\\end{array}\right), \q \q
\hat\n_{m AB}=\left(\begin{array}{ccc}\tilde{\n}_m & 0 & 0 \\0 &
\DD_m & 0 \\0 & 0 & \DD_m \\\end{array}\right)  \ ,\no  \\
&& \label{h} h_{m AB}=\left(\begin{array}{ccc}0
& T^{-kl}_{nm }\f{g}{\Phi^\ast} & T^+_{nm  kl}\f{g}{\Phi}
\\T^+_{ij m r }\f{g}{\Phi} & 0 & 0 \\T^{-ij}_{ m r }\f{g}{\Phi^\ast}
& 0 & 0 \\\end{array}\right)\ ,\eea
\be\label{pis}\Pi_{AB}-\f16 R\ 1_{AB}
=\left(
    \begin{array}{ccc}
     \Pi_{mn}+g_{mn}(\Pi-\f16 R)& - 2g{\DD_r}(\f{1}{\Phi^\ast}T^{-kl}_{rn})
     & - 2g{\DD_r}(\f{1}{\Phi}T^+_{rnkl}) \\
      2T^+_{ijrm}\f{1}{\Phi}{\n_r}g & -\f12D_{ij}^{\ \ kl}+\f{1}{12}1^{kl}_{ij}M
       & \f{\Phi^\ast}{\Phi}T^+_{ij}\cdot T^+_{kl} \\
      2 T^{-ij}_{rm}\f{1}{\Phi^\ast}{\n_r}g & \f{\Phi}{\Phi^\ast}T^{-ij}\cdot
      T^{-kl} & -\f12D^{ij}_{\ \ kl}+\f{1}{12}1^{ij}_{kl}M  \\
    \end{array}
  \right) \no
\ee
Also,  
\bea && \hat{\cal F}_{rs}=[\hat\n_r,\hat\n_s]-\hat\n_{[r}h_{s]}+h_{[r}h_{s]} \cr
&& \ \ \ =\left(\begin{array}{ccc}-{R}_{nm rs }+
 T^{-kl}_{[n[r}T^+_{s]m]kl} & -\tilde\n_{[r}
 (T^{-kl}_{ns]}\f{g}{\Phi^\ast}) &- \tilde\n_{[r}
  (T^+_{ns] kl}\f{g}{\Phi}) \\-\DD_{[r}
 (T^+_{ijs] m}\f{g}{\Phi}) &  \XX  F_{rs}(V) &
 T^+_{ij[r t}T^+_{klts]}\f{\Phi^\ast}{\Phi} \\
 -\DD_{[r} (T^{-ij}_{s] m}\f{g}{\Phi^\ast})
 & T^{-ij}_{[r t}T^{-kl}_{ts ]}\f{\Phi}{\Phi^\ast}
 & \XX F_{rs }(V) \\\end{array}\right)
 \ . \la{y}  \eea
 Applying the  algorithm in \rf{p}   to this operator  we find the total vector-scalar
  sector (1 vector, 6 real scalars)
  contribution to the  logarithmic divergence coefficient
 \bea \la{vsa}  &&
(a_2)_{1,0}=(\f{1}{10}  + { 6 \ov 120} ) C^2
-( \f{31}{180}  +  {6 \ov 360} )E
 +\SS(\tau)  +\f16 F^2_{mn}(V)+\f{1}{48}M^2+\f{1}{8}D_{ij}^{\ \
 kl}D_{kl}^{\ \ ij}
 \no \\
 && \ \
\ \ +(\f23+2)\DD_rT^{-kl}_{rm}\DD_sT^+_{sm kl }+
(\f23+1)R_{mn}T^{-kl}_{mr}T^+_{rnkl}  
 -  \f{\n_n\tau\n_m\bar\tau}{(\Im\tau)^2} T^{-ij}_{nr}T^+_{rmij}
\no \\
&& \ \ \ \ +T^{-ij}_{m a} T^+_{anij}T^{-kl}_{mb}T^+_{bnkl}+
\f23T^{-ij}_{ma}T^+_{anij}T^{-kl}_{mb}T^+_{bnkl}-\f13T^{-ij}_{mn}
T^+_{abij}T^{-kl}_{mn}T^+_{abkl}
\ . \la{jjj} \eea
Here $M$ and $\SS$ were defined in \rf{0},\rf{osi}.


\subsection{Fermionic sector}

 Let us now determine  the fermionic   contribution to \rf{gaa}.
 Squaring the operator in \rf{Hpsi} and putting it into the form \rf{cH}  gives
 \bea
&& {\cal H}_{1/2}= - \left(\begin{array}{cc} \d_i^kP_+ & 0
\\ 0 & \d_k^iP_- \\\end{array}\right)  \DD^2 
+
\left(\begin{array}{cc}     {\cal R}^k_i +e_{ij}e^{jk}&
(\not\!\!\DD e_{ik})
\\(\not\!\!\DD e^{ik}) & {\cal R}^i_k
+e^{ij}e_{jk}\\\end{array} \right)\left(\begin{array}{cc}P_+ &
0 \\0 & P_-
\\\end{array}\right)
\no \\
&& \ \ \ \ \ \ \  \ \ \ \ \  +\ 2\left(\begin{array}{cc}0 & T^{-
}_{ikmr}\g_r\\T^{+ik}_{m r}\g_r & 0
\\\end{array}\right) \left(\begin{array}{cc}P_+ & 0
\\0 & P_- \\\end{array}\right)\DD_m \ , \la{hah} \\
 &&  {\cal R}^k_i\equiv \f14R \d^k_i -\s_{rs}F^k_{i \
rs}(V)+\f12\d^k_i\s_{rs}F_{ rs}(a)\ , \ \ \ \ \ \ \ \  e_{ij}\equiv
\f12E_{ij}+\s\cdot T^-_{ij} \ . \no  \eea
The   corresponding
matrices $\hat{P}$ and $\hat{\cal F}$ in \rf{p} are
\bea  &&
\hat{P}=\left(\begin{array}{cc}    Y^k_i&
(\not\!\!\DD e_{ik})-\DD_m T^{-}_{ik mn }\g_n
\\(\not\!\!\DD e^{ik})-\DD_m  T^{+ik}_{m n}\g_n &  Y_k^i
\\\end{array}\right) \left(\begin{array}{cc}P_+ & 0 \\0 & P_- \\\end{array}\right), \la{fa}\\
&& Y^k_i \equiv  \f{1}{12}R \d^k_i    -\s_{rs}F^k_{i\
rs}(V)+\f12\d^k_i\s_{rs}F_{  rs}(a) +e_{ij}e^{jk}+ T^{-
}_{ijrm}T^{+jk}_{m s}\g_r \g_s \ , \cr &&\hat{\cal
F}_{sr}=\left(\begin{array}{cc} Z^j_{i \ sr} & -\DD_{[s}
T^{-}_{ ik r] m}  \g_m
\\-\DD_{[s} T^{+ik}_{r] m}\g_m & Z^i_{j \ sr}
\\\end{array}\right), \la{fb}\\
&&
Z^j_{i\ sr}  \equiv      \f12R_{sr }^{\ \
 mn}\s_{mn} \d_i^j +F^j_{i \ sr}(V)-\f12\d^j_i
  F_{ sr}(a)+T^{-}_{ik [s m}  T^{+kj}_{r] n}\g_m \g_n\no \ .
\eea
This gives  (for the  number $n_F=\d^i_i$ of 
 Weyl fermions)\footnote{Note the
 following identities
$$
T^{-ik}_{mn} T^{+}_{kj{mn}}+T^-_{jk{mn} }T^{+ki}_{mn}=-\f12\d^i_jT^{-kl}_{mn}
T^{+}_{klmn }\ ,\ \ \ \ \ 
 T^-_{ms} T^+_{sn} = T^-_{ns}T^+_{sm}  \ ,  
 \ \ \ \ \ R_{mn sr}T^-_{ms}T^+_{nr}=-R_{mn}T^-_{ms}T^+_{sn}\ .
$$}
\bea
&&\f12\tr\hat{P}^2=  n_F\big[  \f{1}{72}R^2-
\f14 F^2_{mn}(a)\big]-F^2_{mn}(V)+\f{1}{12}RE_{ij}E^{ij}+\f18 E_{ij}E^{jk}E_{kl}E^{li} \cr
&& \ \  \ \ \ \ \ \ \ \ \ \ \ \
-2\DD_m T^{-}_{klmr}\DD_nT^{+kl}_{nr}+\f12\DD_r E_{kl}\DD_r E^{kl}
\ , \la{pio} \\
&&
\f{1}{12}\tr\hat{\cal F}_{mn}\hat{\cal F}_{mn}= \f{1}{12}\Big[n_F F^2_{mn}(a)+4 F^{2}_{mn}(V)-\f12 n_FR_{sr mn}R^{sr
mn}+8R_{sr }^{\  mn }T^{-}_{kl s m}T^{+kl}_{r n }  \cr
&&
 \ \ \ \ \ \ \ \ \ \ \ \  \  +8(2T^-_{mr  ik}T^{+kj}_{rn}T^-_{ms  jl}T^{+li}_{sn }-
T^-_{mn ik}T^{+kj}_{rs}T^-_{mnjl}T^{+li}_{rs })-8\DD_{s}T^{+ij}_{s m}\DD_{r}T^{-}_{r
mij}\Big].
 \la{tyo}
 \eea
Then finally  we get for the corresponding $a_2$ coefficient
  in \rf{p} (here $n_F=4$   and
we  include   the minus sign
in front of the fermionic contribution in \rf{det})
\bea  &&(a_2)_{1/2}=\f{1}{10} C^2-\f{11}{180}E +\f13 F^2_{mn}(V)+\f13F^2_{mn}(a)\no\\
&& \ \ \ \ \ \ \ \ \ \ \ \ \ \ \ \ \ \ \
 -\f14(\DD_m E_{ij}\DD_m E^{ij}+\f{1}{6}RE_{ij}E^{ij})-\f{1}{16} E_{ij}E^{jk}E_{kl}E^{li}\la{fef}\\
&&\  \  +
\f43\DD_{m}T^{+}_{ijmr}\DD_{n}T^{-ij}_{nr}
   +\f13R_{mn}T^{-kl}_{mr}T^+_{rn kl} +\f13(2T^{-ik}_{mr}T^+_{rnkj }T^{-jl}_{mr}T^+_{rnli }-
T^{-ik}_{mn}T^+_{rskj}T^{-jl}_{mn}T^+_{rsli }) .    \no
\eea
This  expression is obviously $SU(1,1)$ invariant.

\subsection{Final result }

The  total   $\cN=4$   vector multiplet contribution $(a_2)_{\cN=4\, {\rm tot}}$
is given   by the sum of  \rf{jjj}  and \rf{fef}. It
thus starts with $(a_2)_{1,0} + (a_2)_{1/2} =
{1 \ov 4} (C^2 - E) +...= \ha (R^2_{mn} - {1 \ov 3} R^2) + ... $.
The  complete expression may be written as
\bea &&(a_2)_{\cN=4\, {\rm tot}}= \f14  {\cal L}_{_{ {\rm \cN=4\, CSG}} }  \ , \la{al}\\
&&{\cal L}_{_{ {\rm \cN=4\, CSG}} }=
2\big[R_{mn}-\f14\f{\n_{(m}\tau\n_{n)}\bar\tau}{(\Im\tau)^2}
+2T^{-ij}_{ mr}T^+_{ rnij}\big]^2
 -\f23\big[R-\f{\n_m\tau\n_m\bar\tau}{2(\Im\tau)^2}\big]^2\no\\
 && \ \ \ + 2F^i_{j\, mn}(V)F^j_{i\, mn}(V)  +
  \f{1}{(\Im\tau)^2} \big|\nabla^2  \tau +\f{i}{\Im\tau}\n_m\tau\n_m\tau\big|^2 \no \\ 
&& \ \ \ +16\DD_rT^{-ij}_{rm}\DD_sT^+_{sm ij} +\f43\big(2 T^{-ik}_{mr}T^+_{rnkj}T^{-jl}_{ms}T^+_{snli}-
T^{-ij}_{mr}T^+_{rnij}T^{-kl}_{ms}T^+_{snkl}\big)
\no \\
&& \ \ \
-
\DD_r E_{ij}\DD_r E^{ij}-  \f16(R-\f{\n_m\tau\n_m\bar\tau}{2(\Im\tau)^2})E_{ij}E^{ij}
-\f{1}{6} E_{ij}E^{jk}E_{kl}E^{li}
+\f{1}{2}D_{ij}^{\ \ kl}D_{kl}^{\ \ ij}
\la{fin}
\eea
This  should  represent   (up to an  overall factor of 1/4, cf.\rf{cs},\rf{gaa})
 the bosonic part of the full $\cN=4$  conformal supergravity Lagrangian.

This expression passes several checks.
The resulting  action \rf{ccc} is  Weyl invariant; in particular,   all the  fields
have the expected    Weyl-invariant  kinetic terms.
Also,  the truncation  to $\cN=2$ theory  (when $i,j=1,2$) is consistent   with   the
known non-linear action  of $\cN=2$  supergravity \ci{bdw}.

The resulting  CSG Lagrangian    is invariant under the global $SU(1,1)$,
 supporting the  proposal  \ci{bdw}  about  the existence of the full non-linear
$\cN=4$ CSG   action with such symmetry.

The final expression in \rf{fin}  may be rewritten in the manifestly
$SU(1,1)$ invariant form with local $U(1)$  invariance by  replacing the
 $SL(2,R)$ invariants
built out of derivatives of $\tau$  by the corresponding   combinations
involving $\phi_\a$  as in  \rf{aaa}, \rf{3},
or by using the direct  relation between $\tau$ and $\phi_\a$ in \rf{22}
 in the gauge \rf{ga}.
In particular, for the  double-derivative term in \rf{osi},\rf{fin}   one
has $\f{\cD^2\tau \cD^2\bar\tau}{4(\Im\tau)^2}=
(\varepsilon^{\a\b}\phi_\a \DD^2\phi_\b)( \varepsilon_{\g\d}\phi^\g \DD^2\phi^\d)$.

\section{Summary}

The  above computation of  divergent term in 
the $\cN=4$ SYM effective action in  conformal supergravity background 
allowed  us to find   the complete   
$SU(1,1)$ symmetric action of $\cN=4$ conformal supergravity in the 
bosonic sector.
We used  that the 
divergent part of the effective action is local, preserves all the 
symmetries of the underlying classically superconformal 
theory and  starts 
with the Weyl tensor squared  term.

The fermionic part of the $\cN=4$ conformal
supergravity action can be found by the same method. Indeed, 
the   $\N=4$  SYM -- CSG coupling   given in \ci{roo}   contains all the 
required fermionic terms. This  is  still  straightforward   but 
 technically  more involved.

\

{\bf Note added:} 
The   computation  of the  bosonic terms in the   $\cN=4$ conformal
supergravity action  viewed  as an     induced action 
reported in this paper  missed some  of the relevant  terms when including the fermionic sector contributions in  section 3.2.
The complete expression for the $\cN=4$ conformal
supergravity action was  found later   \ci{Ciceri:2015qpa} by a  direct 
 method   based on  supersymmetry    (see  discussion in  section 5  in  \ci{Ciceri:2015qpa}). 
We thank   F. Ciceri, B. de Wit   and B. Sahoo  for pointing this out.

\section*{Acknowledgments }
A.A.T. would like to thank R. Kallosh  and R. Roiban for discussions
of related models. The work of I.L.B. and N.G.P. was partially
supported by RFBR grant, project No 12-02-00121 and by a grant for
LRSS, project No 224.2012.2.  Also, I.L.B. acknowledges the support
of  the RFBR-Ukraine grant, project No 11-02-90445 and DFG grant,
project No LE 838/12-1. N.G.P.\ acknowledges the support of the RFBR
grant, project No 11-02-00242. The work of A.A.T. was supported by
the STFC grant ST/J000353/1 and by the ERC Advanced grant No.290456.

\bigskip

\end{document}